\documentclass[%
reprint,
superscriptaddress,
 amsmath,amssymb,
 aps,prx
]{revtex4-2}

\usepackage{graphicx}
\usepackage{dcolumn}
\usepackage{bm}
\usepackage[dvipsnames]{xcolor}

\usepackage[colorlinks=true,citecolor=blue]{hyperref}
\hypersetup{colorlinks=true,citecolor=blue,linkcolor=blue,urlcolor=blue}
\renewcommand\labelenumi{(\roman{enumi})}
\renewcommand\theenumi\labelenumi
\usepackage{enumitem}
\usepackage[normalem]{ulem}

\begin{document}


\title{Non-adiabatic polariton condensation in annular optical traps}
\author{I. Chestnov}
\affiliation{School of Physics and Engineering, ITMO University, Kronverksky Pr. 49, bldg. A, St. Petersburg, 197101, Russia}
\author{E. Cherotchenko}
\affiliation{Ioffe Institute, ul. Polytekhnicheskaya 26, St. Petersburg, 194021, Russian Federation}
\author{A. Nalitov}
\affiliation{Faculty of Science and Engineering, University of Wolverhampton, Wulfruna Street, Wolverhampton WV1 1LY, United Kingdom}
\affiliation{Moscow Institute of Physics and Technology, Institutskiy per.,~9, Dolgoprudnyi, Moscow Region, Russia 141701}

\begin{abstract}
In analogy with superfluidity, nonequilibrium polariton condensation can be phenomenologically described in terms of separated condensed and normal fractions.
Under the assumption of significantly shorter characteristic timescale, the incoherent part is frequently traced out adiabatically.
In this work, we stress the importance of accounting for the time-resolved coupling between the condensed and normal fractions.
Even in the case of a significant mismatch in the evolution rates, the coupling with the normal fraction drastically alters the condensate dynamics and leads to a variety of previously overlooked phases in a confined configuration.
Focusing on the case of annular optically induced traps for polaritons, we elaborate on the non-adiabatic model of polariton condensation.
Using the two-mode approximation, accounting for a couple of quantized polariton vortices with opposite angular momenta, we identify the range of validity for the adiabatic elimination of the incoherent reservoir.
Beyond this range, the non-adiabatic interaction with a circularly symmetric trapping reservoir supports multistability, limit cycle dynamics, and formation of a neutral equilibrium standing wave phase featuring spontaneous breaking of the continuous radial symmetry.
In the presence of a weak trap asymmetry, the non-adiabaticity prevents formation of giant polariton vortices.
We argue that our detailed description gives interpretation to the dominance of the standing wave phase over persistent vortex phase in experimental observations.
\end{abstract}

\maketitle

\section{introduction}

Optically induced traps represent a versatile platform for creating and manipulating non-equilibrium bosonic condensates of exciton polaritons \cite{Ohadi2016,Schneider2017,Alyatkin2020}. 
Recently developed techniques of all-optical control over polariton condensates allow for various potential applications in lattice simulators \cite{Ohadi2017,Berloff2017,Alyatkin2021}, topological photonics \cite{Sigurdsson2019,Pickup2020,Harrison2023}, and qubit engineering \cite{Xue2021,Kavokin2022,Barrat2023}. 
The trapping potentials of this type are typically formed by spatially sculptured nonresonant optical pump creating incoherent excitons \cite{KavokinMicrocavities,Carusotto2013}. 
The resulting hot exciton gas simultaneously confines the coherent condensate due to interparticle repulsion and sustains its density due to stimulated scattering from the incoherent fraction compensating radiative losses. 
Because of this driven-dissipative nature, polariton condensates can occupy excited states of the trap, which in combination with strong exciton-exciton interactions gives rise to nonlinear bifurcations and multistability in spin \cite{Ohadi2015,Askitopoulos2016a} or orbital \cite{Askitopoulos2018} degrees of freedom. 

The mean field model of polariton condensation  typically takes advantage of a reduced description of the hot exciton gas treating it as a completely uncorrelated \textit{reservoir} \cite{Wouters2007a} which is fully characterized by its density distribution. 
The further simplification of this model employs \textit{adiabatic} approximation by assuming a rapidly adapting reservoir, which is finally eliminated. 
This leads to the open-dissipative Gross-Pitaevskii equation (ODGPE) for the condensate order parameter \cite{Keeling2008}.

The validity of adiabatic approximation relies on several conditions, one of which requires a low ratio of the condensate and reservoir decay rates \cite{bobrovska2015}. 
At the same time, polariton lifetime is mainly governed by the cavity quality factor \cite{KavokinMicrocavities} and doesn't exceed a few hundreds of picoseconds \cite{Steger2013}.
Yet, the reservoir lifetime is determined by the much slower non-radiative exciton decay \cite{baboux2018} which occurs on the sub-nanosecond timescale \cite{Szczytko2004}.
Nevertheless, the ODGPE model with adiabatically excluded reservoir proved very efficient for describing experimentally observed phenomena with incoherently pumped polariton condensates.
However, certain effects, such as reservoir depletion \cite{Estrecho2018} and condensate instability due to effective reservoir-mediated attraction  \cite{Wouters2007a,Smirnov2014,Liew2015}, can only be described in the full condensate+reservoir model and are therefore inherently non-adiabatic.

An important application of optical traps is creation of the annular geometry which allows formation of whirling polariton condensates, exhibiting persistent rotating currents and emitting coherent optical vortices.
However, in the absence of explicit chiral symmetry breaking such as rotation \cite{gnusov2023,Yulin2023a} or gain chirality \cite{Sedov2021PRR}, annular traps were shown to exhibit polariton condensation in the combinations of nonzero orbital momentum states with no net vorticity \cite{Dreismann2014b,Sun2018}. 
In this case, the condensate resembles the standing-wave superposition of whispering gallery modes or counter-rotating giant vortices. 

The mean-field ODGPE description attributed such a non-chiral condensation to the external breaking of continuous rotational symmetry due to disorder or pumping asymmetry which couples counter-rotating polaritons \cite{Nalitov2019}. 
Even then, the ODGPE predicts destabilization of the standing-wave condensate with increasing pump power. 
The condensate is expected to transform into a giant vortex with a high angular quantum number due to spontaneous chiral symmetry breaking triggered by the nonlinear interactions.
However, such spontaneous formation of vortices with stochastic rotation direction, randomly selected during the condensate formation stage, was only observed in small traps supporting vortex angular number $l=1$ \cite{Cookson2021}.

In this work, we demonstrate that the observed dominance of non-chiral states over vortices may be interpreted as a manifestation of the non-adiabatic coupling with an annular  reservoir of incoherent excitons (hereafter referred to as reservoir for brevity).
Accounting for two oppositely rotating polariton modes coupled to three relevant angular harmonics of the reservoir, we describe the reported non-chiral condensation by a mechanism akin to the reservoir-mediated self-trapping effect~\cite{Wouters2007a}. 
This result can be analytically obtained once circular symmetry of the trap is respected and proven numerically in the case of a reduced symmetry.
We further explore the respective four-dimensional parameter space of the problem and, in addition to vortex-like and non-chiral condensates, describe peculiar dynamical phases with self-pulsating density angular distribution and vorticity \cite{Barrat2023}.

This paper is organized as follows.
We develop the non-adiabatic mean-field model and discuss its experimentally relevant parameters in Sec.~\ref{sec:2modes}.
Stability of analytically tractable stationary solutions of the non-adiabatic model is discussed in Sec.~\ref{Sec:StatSol}.
Modification of the model due to asymmetry of the incoherent pump and its impact on the stability of both non-chiral and stochastic vortex condensates is described in Sec.~\ref{sec:defect}.
The conditions allowing adiabatic elimination of the reservoir, as well as the differences in predictions of the full model and its adiabatic approximation are discussed in Sec.~\ref{sec:adiabatic}.
In the concluding section we provide a list of possible implications of the non-adiabatic effects in the interpretation of recent experiments with ring-shaped polariton condensates.

\section{non-adiabatic two-mode model} \label{sec:2modes}

\begin{figure}
    \centering
    \includegraphics[width = \linewidth]{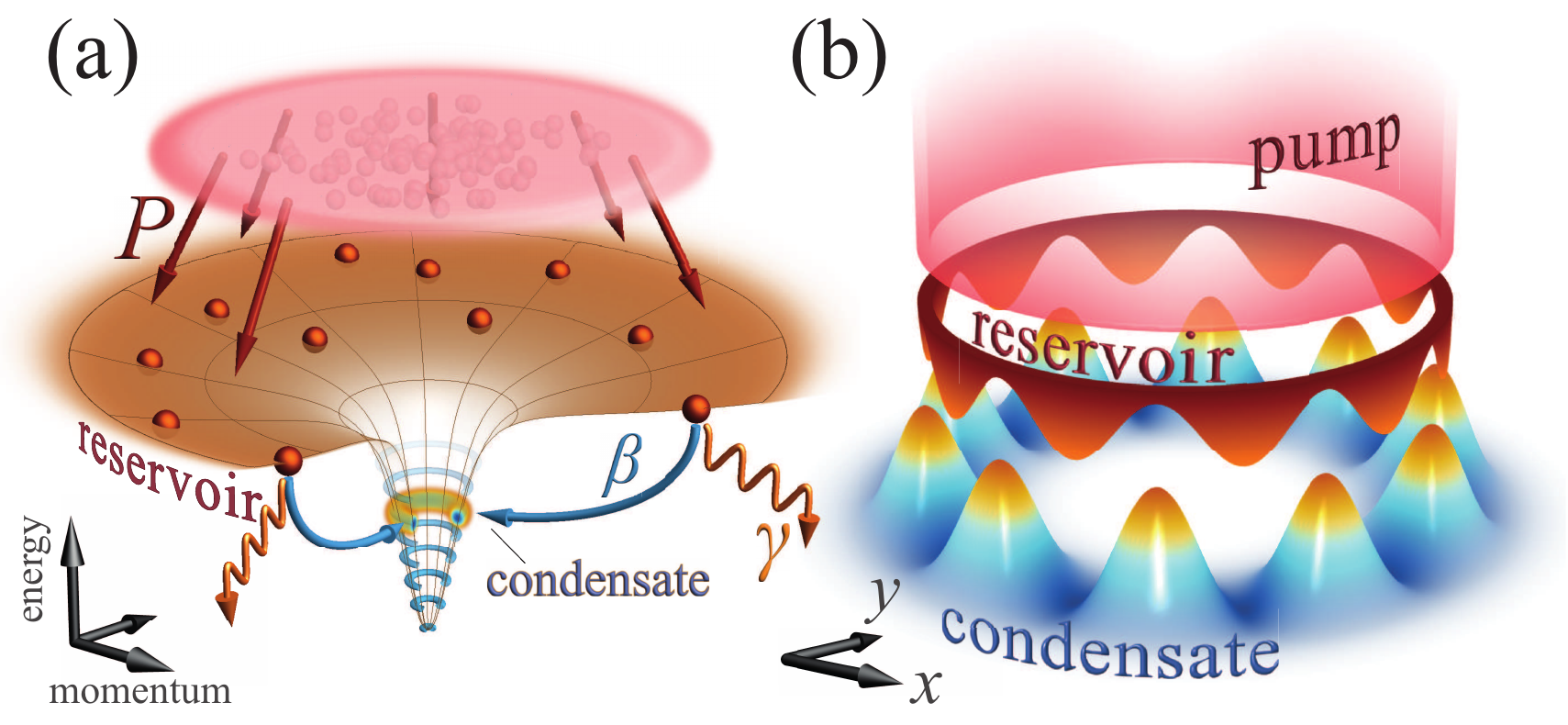}
    \caption{(a)  Momentum-space representation of the non-resonantly excited polariton condensate. The reservoir of high-momentum states of the lower polariton dispersion branch are created from the continuum of free carriers (dim red cloud) generated under high-energy laser radiation. Due to phonon-mediated cooling, these exciton-like particles (red beads) accumulate near the region of the dispersion inflection from where they can be scattered (blue arrows) to the discrete set of leaky polariton modes (blue-shaded levels) that arise from the spatial quantization in the trap. The mode with the fastest net gain rate (not necessarily the lowest in energy) accumulates a microscopically large number of particles thereby manifesting formation of the condensate (glowing yellow level). The ODGPE model \eqref{eq:FullModel} describes dynamics of the incoherent reservoir by the net scattering rate $\beta$ and the effective reservoir relaxation $\gamma$ and filling $P$ rates. (b)  A sketch of the burner-shaped polariton condensate created with a ring-shaped optical pump. The density of the ridged reservoir is shown upside down.}
    \label{fig:sketch}
\end{figure}

We begin with the standard semi-phenomenological description of the non-resonantly excited nonequilibrium bosonic condensate of microcavity polaritons \cite{Wouters2009,Haug2014} sketched in Fig.~\ref{fig:sketch}(a):
\begin{subequations}\label{eq:FullModel}
    \begin{eqnarray}
        \label{eq:GPE}
    i\hbar \partial_t \Psi &=& \left[ - {\hbar^2 \Delta \over 2m} 
    + n {\alpha + i \beta \over 2} - i \hbar {\Gamma \over 2} + \alpha_1 |\Psi|^2 \right]\Psi,\\
    \partial_t n &=& P - \left( \gamma + {\beta |\Psi|^2 \over \hbar} \right)n. \label{eq:reservoir}
    \end{eqnarray}
\end{subequations}
Here $\Psi$ and $n$ are the condensate wave function and the reservoir density, $m>0$ is a polariton effective mass, $\Delta$ stands for the two-dimensional Laplace operator, $\alpha>0$ governs the strength of the condensate repulsive interaction with the reservoir, while $\alpha_1>0$ is a parameter of polariton-polariton repulsion within the condensate, $\beta>0$ determines the rate of stimulated scattering from the reservoir into the condensate, $\Gamma>0$ and $\gamma>0$ are the polariton and exciton decay rates, respectively.
We neglect the thickness of the ring-shaped pump profile and assume its form as $P(r) = P_0 \delta(r-R)/R$ with $R$ being the radius of the annular trap and $r$ -- the radial coordinate. 
A sketch of the real-space polariton density in the considered system is depicted in Fig.~\ref{fig:sketch}(b). 

\subsection{Linear limit}\label{subsec:IIA}

We first address the linear case neglecting nonlinear terms in Eqs.~\eqref{eq:FullModel}, which is valid below and in the vicinity of the condensation threshold. 
The complex spectrum of the resulted linear problem allows  determination of the threshold pumping strength together with the properties of condensate state, namely, its angular momentum $l$ and single-polariton energy $E^l_\text{th}$.

Because of the rotational symmetry of the problem, it is natural to search for the condensate eigenstates in the form of vortices parameterized with an integer phase winding number $l$. 
With exciton diffusion neglected, the reservoir profile follows that of the pump:
\begin{equation}
    n(r) = {P_0 \over \gamma} {\delta(r-R) \over R}.
\end{equation}
In this case polariton modes can be piecewise defined inside and outside the circle of radius $R$:
\begin{equation} \label{eq:resonance}
    \Psi_l(r,\varphi) = \exp(il\varphi - i\omega_0 t) \times \left\lbrace \begin{matrix}
        A J_l (\varkappa r), & r<R \\
        B H_l (\varkappa r), & r>R
    \end{matrix} \right.,
\end{equation}
where $J_l$ and $H_l$ are the Bessel and Hankel functions of the first kind, $\varkappa = \sqrt{2m(\omega_0 + i \Gamma/2 )/\hbar}$ with the principal square root value, $\varphi$ is the polar angle. 
The coefficients $A$ and $B$ are determined from the wave function continuity  at $r=R$ and the normalization condition $\int |\Psi_l|^2d^2r=1$. %

Let us define a critical pump amplitude $P_\textrm{th}^l$, above which the $l$-th azimuthal mode is amplified. Exactly at $P_0 = P_\textrm{th}^l$ the single-polariton energy $ \hbar \omega_0 = E_\textrm{th}^l$ is purely real. 
This value can be found from the eigenvalue problem which reduces to 
\begin{equation}\label{Eq.EigenProblem}
    \frac{J_l(\varkappa R)\left[ l H_l(\varkappa R) - \varkappa R H_l(\varkappa R) \right]}{H_l(\varkappa R) \left[ l J_l(\varkappa R) - \varkappa R J_{l+1}(\varkappa R) + c J_l(\varkappa R) \right]} = 1,
\end{equation}
with $c = m(\alpha + i \beta) {P_\textrm{th}^l} / (\hbar^2 \gamma)$.

If the above-threshold condition $P_0 > P_\textrm{th}^l$ holds for several $l$, these modes compete with each other during the onset of the condensation. 
Typically, the winner depends on the weak fluctuating occupation of the modes at the moment of time when the pump is switched on.
However, if the pump intensity increases slowly at the condensate formation time scale (about few picoseconds), the condensation is expected to occur in the mode with the lowest threshold $P_\textrm{th}^l$. 
Note that the perfectly circular pump implies the possibility of spontaneous chiral symmetry breaking as all excited modes $|l| \geq 1$ are twice degenerate, so that $P_\textrm{th}^l=P_\textrm{th}^{-l}$ and $E_\textrm{th}^l = E_\textrm{th}^{-l}$.

\begin{figure} 
    \centering
    \includegraphics[width=1\linewidth]{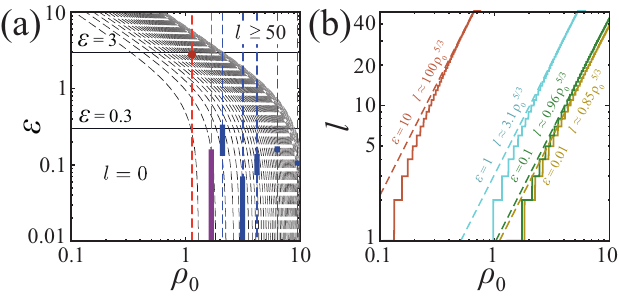}
    \caption{
   Predictions of the linear model based on numerical solution of  the eigenvalue problem \eqref{Eq.EigenProblem}. 
    (a) The linear phase diagram on the parameter space spanned by the normalized trap size $\rho_0 = R/R_0$ and the ratio $\varepsilon=\alpha/\beta$. Each grey dashed line corresponds to the angular number increment by 1. The leftmost region corresponds to ground state condensation with $l=0$, transitions at $l\geq50$ are not shown. The dashed vertical lines correspond to the normalized radii of the trapping potentials realized in Refs.~\cite{Dreismann2014b} (blue, $l=3,5,10,22,40$), \cite{Sun2018} (red, $l=14$), and \cite{Askitopoulos2018} (magenta, $l=1$). Solid thick regions highlight the allowed range of the $\varepsilon$ parameter, corresponding to experimentally observed values of the angular quantum number $l$.
    (b) The cascade of the condensate angular number transitions which occur with increasing trap size $\rho_0$. The solid lines correspond to the angular number $l$ minimizing $P^l_\text{th}$. The dashed lines stand for the asymptotic fitting for $l\gg1$ for $\varepsilon \equiv \alpha/\beta=0.01$, $0.1$, $1$, $10$ (yellow, green, cyan, orange).}
    \label{fig:linear_model}
\end{figure}

Introducing the characteristic polariton attenuation length $R_0=\sqrt{\hbar/(m\Gamma)}$, the parameter space of the linear model can be reduced to only the dimensionless trap size $\rho_0=R/R_0$ and the ratio $\varepsilon = \alpha/\beta$ between the real and imaginary parts of the condensate-reservoir interaction.
Numerically solving the eigenvalue problem \eqref{Eq.EigenProblem} and minimizing obtained values of $P^l_\text{th}$ one obtains the condensate angular number $l(\rho_0,\varepsilon)$ \cite{Cherotchenko2021}.
This dependence presents a ladder of incremental transitions $l\rightarrow l+1$ from $l=0$ (at low values of both $\rho_0$ and $\varepsilon$), shown with dashed lines in Fig.~\ref{fig:linear_model}(a).
Fig.~\ref{fig:linear_model}(b) shows that the dependence $l(\rho_0)$ may be approximated as $l\sim a(\varepsilon) \rho_0^{5/3}$ for sufficiently high angular numbers $l\sim50$ in a broad range of $\varepsilon$ values, in contradiction to previously reported quadratic dependence \cite{Dreismann2014b}.

Comparing the diagram in Fig.~\ref{fig:linear_model}(a) with the experimental observations reported in Refs.~\cite{Dreismann2014b,Askitopoulos2018,Sun2018}, we estimate the ratio of the condensate-reservoir interaction parameters $\varepsilon$  to be below 1 ($\alpha/\beta\sim0.1$) in the former reports \cite{Dreismann2014b,Askitopoulos2018} [with $\Gamma \approx (10\,\text{ps})^{-1}$, $m\approx5\times10^{-5}m_{\rm e}$, $R_0 \approx 5\,\mu\text{m}$] and above 1 ($\alpha/\beta\approx3$) in the latter setup \cite{Sun2018} [with $\Gamma \approx (135\,\text{ps})^{-1}$, $m\approx5\times10^{-5}m_{\rm e}$, $R_0 \approx 18\,\mu\text{m}$].
The corresponding ranges of the estimated $\alpha/\beta$-ratio are highlighted with thick color lines and the red dot in Fig.~\ref{fig:linear_model}(a).
In the following sections, we reveal a qualitatively different behaviour of the condensate in these two regimes.

\subsection{Nonlinear extension} \label{subsec:nonlinear_model}

To describe the condensate dynamics beyond the condensation threshold, we extend our description accounting for the nonlinear terms in Eqs.~\eqref{eq:FullModel}.
However, we focus on the regime of weak interactions, including them perturbatively and using the  linear model spectrum as an unperturbed seed.
This proves minimally sufficient to analytically describe such effects as spontaneous formation of chiral vortex condensates, multistability, and dynamic regimes of alternating vorticity.

Using the radial symmetry of the system, we employ a two-mode approximation valid in the weak interaction limit, where the interaction energy is small compared to the energy distance to the nearest levels of the linear spectrum.
We thus assume that the degenerate doublet of $\pm l$ states \eqref{eq:resonance} is decoupled from all other states, and the condensate wave function reads
\begin{equation} \label{eq:WF}
    \Psi = \psi_+ \Psi_{+l} + \psi_- \Psi_{-l},
\end{equation}
where $\psi_\pm$ are complex numbers. 
Expression \eqref{eq:WF} enables azimuthal dependence of the reservoir density with the angular harmonics being twice the condensate orbital number $l$:
\begin{equation}\label{eq:Res_Den_Definition}
    n(r,\varphi) = \left[ n_\textrm{th} + n_0 +n_1 \cos(2l\varphi) + n_2 \sin(2l\varphi)\right] {\delta(r-R)\over R},
\end{equation}
where $n_\textrm{th} = P_\textrm{th}^{l}/\gamma$ and $n_0$, $n_1$, and $n_2$ are the three real numbers defining the reservoir density azimuthal distribution.

Let us now introduce the three-dimensional vector $\bm{s}=\psi^\dagger \bm{\sigma} \psi/2$ analogous to the classical spin, with $\psi = [\psi_+, \psi_-]^\intercal$ and $\bm{\sigma}$ being the Pauli vector.
The magnitude $s$ governs the condensate population while the direction $\bm{s}/s$ describes its rotational degree of freedom.
In particular, the $\bm{s}$-vectors belonging to the $xy$ plane correspond to non-rotating condensates with $2l$ nodes in the angular density distribution.
In what follows we refer to these standing-wave condensates as ``petals'' following Ref.~\cite{Dreismann2014b} or ``burner'' due to its visual similarity with the flame of a gas range burner, -- see Fig.~\ref{fig:sketch}(b).
In contrast, vectors close to the $z$ axis describe the vortex states with the rotation direction determined by the sign of $s_z$.

We therefore proceed with projecting Eq.~\eqref{eq:FullModel}  into the spin space with the use of Eq.~\eqref{eq:Res_Den_Definition} for the reservoir density components, see Appendix~\ref{AppA}. 
Then, using the dimensionless time $\tau = \gamma t$ and introducing $\bm{S} = b\bm{s}/\gamma$, $N_j = 2\pi b n_j/\gamma$, where $b=\beta|\Psi_l(R)|^2 /\hbar$ stems from the condensate overlap with the $\delta$-ring reservoir, we obtain the dimensionless two-mode model of non-adiabatic polariton condensation:
\begin{subequations}\label{eq:nonAd0Dmodel}
    \begin{eqnarray}
     \dot{S_x} &=& N_0S_x + {1 \over 2} N_1 S + {\varepsilon \over 2} N_2 S_z  +\xi S_yS_z, \label{eq:dSx}\\
    \dot{S_y} &=& N_0S_y + {1 \over 2}N_2 S - {\varepsilon \over 2} N_1 S_z -\xi S_xS_z,\label{eq:dSy}\\
    \dot{S_z} &=& N_0S_z + {\varepsilon \over 2} (N_1 S_y -N_2 S_x),\label{eq:dSz}\\
    \dot{N_0} &=& W - (1+2S)N_0 - 2SN_\textrm{th} -S_xN_1-S_yN_2,\label{eq:dN0}\\
    \dot{N_1} &=& -(1+2S)N_1 -2S_x(N_\textrm{th}+N_0),\label{eq:dN1}\\ 
    \dot{N_2} &=& -(1+2S)N_2 -2S_y(N_\textrm{th}+N_0).\label{eq:dN2} 
    \end{eqnarray}
\end{subequations}
Here $\varepsilon=\alpha/\beta$  and  $\xi = a/b$ with $a = 4\pi \alpha_1 \int_0^\infty |\Psi_l|^4 r dr / \hbar$ accounting for condensate self-interactions. 
The pump variation from the threshold value is quantified by $W = 2\pi b(P-P_\textrm{th}^l)/\gamma^2$. It is noteworthy that in the adopted notations, the normalized threshold  density of the reservoir $N_\textrm{th} = \Gamma/\gamma$ is equal to the ratio between the reservoir and the condensate lifetimes, which is typically used as a main criterion of the adiabatic elimination of the reservoir \cite{bobrovska2015}. 

\begin{figure} 
    \centering
    \includegraphics[width=1\linewidth]{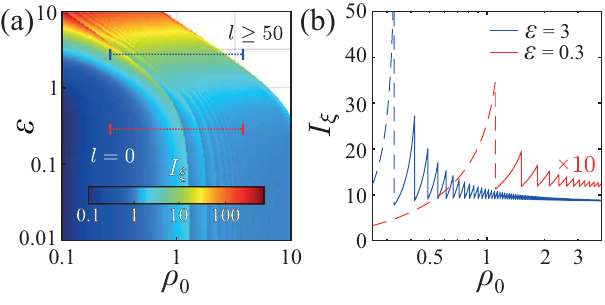}
    \caption{(a) Dimensionless form-factor $I_\xi$ given by Eq.~\eqref{eq:Ixi}.
    (b) Cross-sections at $\varepsilon=3.0$ and $0.3$, shown with dotted lines in (a).
    The values of $I_\xi$ for $\varepsilon=0.3$ are multiplied by 10 for visibility.
    In the range of trap sizes, corresponding to condensation at the ground state $l=0$, where the two-mode model is inapplicable, $I_\xi$ is plotted with the dashed lines.}
    \label{fig:linear_model_b}
\end{figure}

Parameter $\xi$ in Eqs.~\eqref{eq:nonAd0Dmodel} quantifies the strength   of the condensate self-repulsion accounting for the condensate density profile and its overlap with the reservoir. 
As it is shown in Appendix A, one can introduce the dimensionless form-factor $I_\xi$ which quantifies the condensate wave function localization:
\begin{equation} \label{eq:Ixi}
    I_\xi = \frac{2}{p_\textrm{th}^l} \left[ \int_0^{\rho_0} \left|\frac{J_l(\kappa\rho)}{J_l(\kappa\rho_0)}\right|^4 \rho d\rho + \int_{\rho_0}^{\infty} \left|\frac{H_l(\kappa\rho)}{H_l(\kappa\rho_0)}\right|^4 \rho d\rho   \right],
\end{equation}
such as the $\xi$-parameter is expressed as $\xi \equiv \alpha_1 I_\xi / \beta$.  Here $\kappa = \sqrt{2\omega_0/\Gamma + i}$ with the principal square root value.

The numerically computed parameter $I_\xi$ extracted from the solution of the eigenvalue problem \eqref{Eq.EigenProblem} is shown in Fig.~\ref{fig:linear_model_b}.
Note that for any fixed  $\varepsilon$,  $I_\xi$ reaches its maximal value at a small trap size corresponding to the $l=1$ condensate.
Surprisingly, at larger trap sizes, $I_\xi$ saturates at a certain level whose value increases with $\varepsilon$, see Fig.~\ref{fig:linear_model_b}(b).

\section{Stationary states}\label{Sec:StatSol}

The non-adiabatic two-mode model \eqref{eq:nonAd0Dmodel} has two types of stationary solutions: the vortex and the burner-like condensates. 
In particular, there are two symmetric vortex states on the $z$ axis in the $\bm{s}$-vector space. They are characterized by $S_z = \pm S$ with $S =  W/(2N_\textrm{th})$, and vanishing reservoir components $N_0=N_1=N_2=0$ according to Eqs.~\eqref{eq:dN0}--\eqref{eq:dN2}.
In addition, due to the circular symmetry of the problem,  there is a continuum of burner-type solutions on the $(x,y)$-plane which corresponds to the ambiguity in the azimuthal orientation of polariton density petals.
These states are given by  $N_{1,2} = - 2 N_0 S_{x,y}/S$ and $S = (W-N_0)/(2N_\textrm{th})$ with 
\begin{equation}
    N_0 =\frac{1}{2}\left[3N_\textrm{th}+W - \sqrt{(3N_\textrm{th}+W)^2 - 4 N_\textrm{th}W}\right].
\end{equation}

It is crucial to stress that only those state which are stable against weak perturbations can be observed experimentally.
Dynamical stability of both vortex and burner states is governed by the eigenvalue spectrum of the corresponding Jacobi matrices.
Although this problem is analytically tractable, the involved expressions are quite cumbersome and are therefore omitted here. 
Note that broken continuous radial symmetry implies emergence of a Goldstone mode in the corresponding excitation spectrum \cite{Goldstone1961}, responsible for neutral stability of burner states with respect to azimuthal orientation of petals.

To limit the dimensionality of the parameter space, we consider two experimentally relevant values of $\varepsilon$-parameter, namely, the case of a strong ($\varepsilon=3$) and a weak ($\varepsilon=0.3$) repulsive interaction between the condensate and reservoir, see Fig.~\ref{fig:linear_model}a.
In the following we show that the behaviour of the condensate is qualitatively different in these two regimes.
In particular, burners are only stable at $\varepsilon>1$, while in the opposite case $\varepsilon<1$ the system exhibits peculiar nonlinear periodic behaviour, where the condensate alternates its rotation direction in a strongly anharmonic manner.

\begin{figure}
    \includegraphics[width=\linewidth]{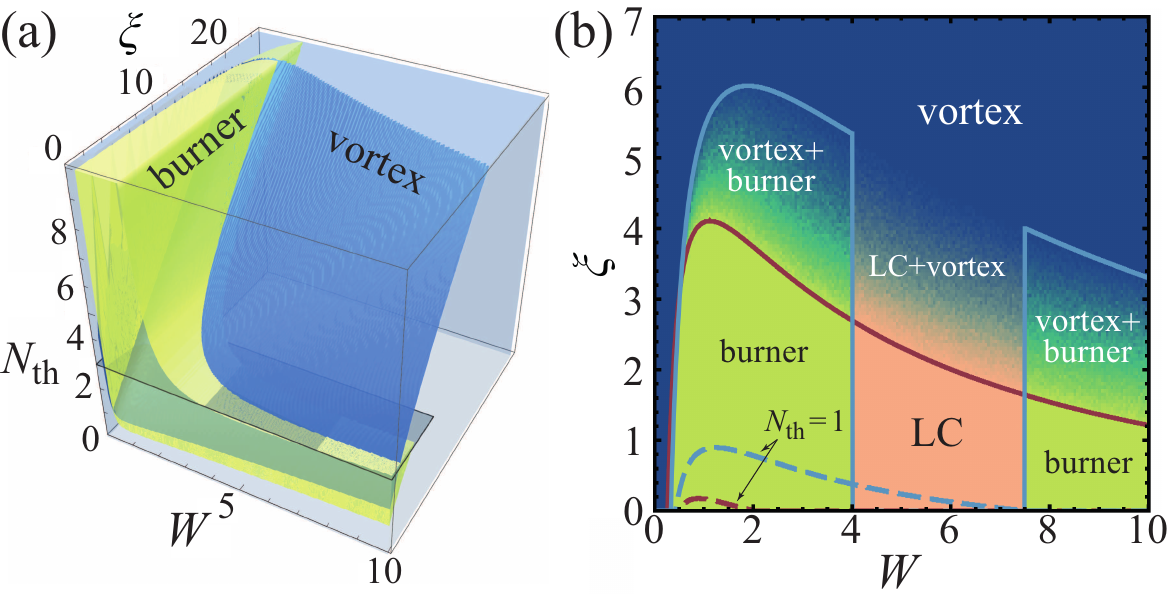}
    \caption{
    (a) Stability diagram at $\varepsilon=3$. Overlapping yellow and blue regions correspond to stability domains of the burner-shaped and vortex condensates respectively. 
    (b) Stability diagram horizontal cross-sections at $N_\text{th}=3$ (solid lines) and $N_\text{th}=1$ (dashed lines). Realization probabilities of petals (yellow), vortices (blue), and limit cycles (LC, peach tone) are shown with color for the case $N_\text{th}=3$.}
    \label{fig:Stability}
\end{figure}

Let us first address the strong-repulsion case $\varepsilon=3$, where the elastic condensate-reservoir scattering dominates over inelastic processes.
The corresponding stability diagram is demonstrated in Fig.~\ref{fig:Stability}.
In particular, stability regions of the vortex and the burner solutions are shown in the 3D space of parameters $N_\text{th}$, $\xi$, and $W$.
Figure~\ref{fig:Stability}(b) illustrates the horizontal cross-sections of this diagram in detail.
The solid lines show the boundaries of the stability domains for burners (blue) and vortices (red) at $N_\text{th}=3$ while the dashed lines correspond to $N_\text{th}=1$.
The vertical segments of the burners stability boundary correspond to a Hopf bifurcation, where fixed-point solutions on the $(x,y)$-plane continuously evolve into closed orbits representing stable limit cycles (LCs).

To illustrate multistability among vortices, burners and LCs, we numerically integrate Eqs.~\eqref{eq:nonAd0Dmodel} with randomly distributed initial conditions.  
The probabilities of attraction to either of stable states are shown with color in Fig.~\ref{fig:Stability}(b).
Within the region of multistability, the volume of the relative basin of attraction is gradually transferred from the vortex to either petal or LC. 
Note that within the whole parameter space, the rotation direction of the observed vortex solutions is stochastic in a sense that it depends on the initial spin orientation seeded at the condensate formation stage.

It is crucial that in the absence of external radial symmetry breaking, the ODGPE-based model predicts an ubiquitous instability of the burner-type condensate \cite{Askitopoulos2018,Nalitov2019}.
Indeed, under the assumption of a quickly adopting reservoir, the system is reduced to the model of two modes coupled via a common depletable gain source, see Sec.~\ref{sec:adiabatic} for the details. 
The eliminated reservoir leaves a strongly asymmetric gain-saturation mechanism \cite{Yulin2023b} which enters as $\dot{\psi}_{\pm} \propto -\left(|\psi_{\pm}|^2 + 2 |\psi_{\mp}|^2\right) \psi_{\pm}$ in the corresponding mode evolution equations.
Under the assumption of a small condensate occupancy, the corresponding equation for the $\bm{S}$-vector reads:
\begin{equation}\label{eq:2mResFreeModel}
    \dot{\bm{S}} = W\bm{S} - \left(\bm{g} \cdot \bm{S}\right) N_{\rm th} S   + \left[\bm{S} \times \bm{B} \right],
\end{equation}
where $ \bm{B} = \left(0,0,\left[ \xi - N_{\rm th} \varepsilon \right] S_z \right)^\intercal$ and $\bm{g} = \left(3,3,2\right)^\intercal$ describes the gain depletion anisotropy.
An equal-amplitude configuration corresponding to the petalled solution is unstable against weak population imbalance. 
In particular, a small addition to the population of mode $l$ reduces gain of the opposite mode $-l$ to a lesser extent than its own. 
The resulting gain imbalance grows and eventually leads to the vortex formation. This clearly contradicts experimental findings \cite{Sun2018,Dreismann2014b} that reported existence of the stable burner-like condensates.

The emergent stability of burners in the present non-adiabatic model \eqref{eq:nonAd0Dmodel} may be interpreted  via so-called hole burning effect \cite{Estrecho2018,Wouters2007a,Ma2015}. 
In this picture, the petalled condensate locally reduces reservoir density thus digging an azimuthally periodic potential grating for itself which prevents vortex formation as sketched in Fig.~\ref{fig:sketch}(b).
However, sufficiently strong self-repulsion quantified by the parameter $\xi$ counters reservoir-mediated trapping effect and therefore favours vortex formation, see Fig.~\ref{fig:Stability}. 

\begin{figure}
    \centering
    \includegraphics[width=\linewidth]{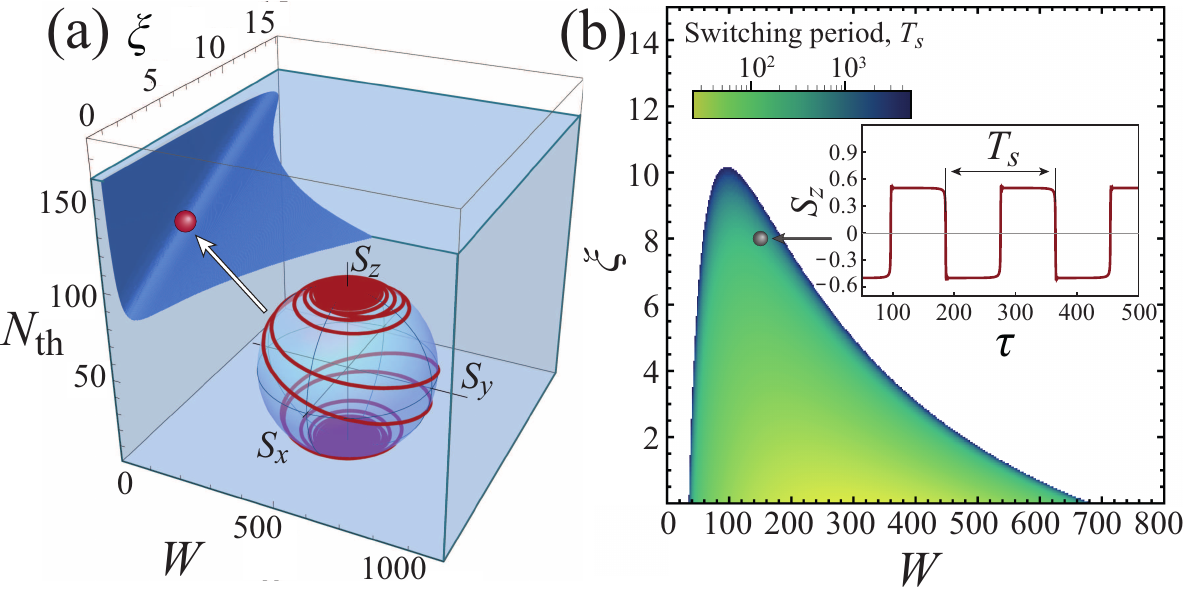}
    \caption{(a) Stability diagram at $\varepsilon = 0.3$. Far above the threshold, vortex is a single stable state. The blue surface indicates the vortex stability boundary. Above it, polariton condensate exhibits regular switching between vortex states. The inset demonstrates a corresponding trajectory in the $\bm{S}$-vector space. (b) The switching period $T_s$ on the $\left(W,\xi\right)$ parameter plane at $N_{\rm th}=150$. The inset demonstrates time dependence of $S_z$ at $W=150$ and $\xi=8$.  }
    \label{Fig3}
\end{figure}

In addition, at the weak condensate-reservoir repulsion, $\varepsilon<1$, the reservoir-mediated potential is unable to stabilize burner solutions which unavoidably transform to the vortex states.  
While burners are always unstable at $\varepsilon<1$, the vortex states lose their stability at sufficiently high values of $N_\text{th}\equiv \Gamma/\gamma$.
The corresponding 3D stability diagram for the case of $\varepsilon=0.3$ is shown in Fig.~\ref{Fig3}(a). 
Beyond the vortex stability domain, the numerical solution of Eqs.~\eqref{eq:nonAd0Dmodel} reveals establishment of a limit cycle behaviour, illustrated in the inset in Fig.~\ref{Fig3}(a). In this regime, the condensate remains close to the vortex state most of the time, while its vorticity 
periodically changes direction by means of rapid jumps, see the inset in Fig.~\ref{Fig3}(b).
The period of vorticity switching $T_s$ shown on the $(W,\xi)$-plane in Fig.~\ref{Fig3}(b) reaches $10^3$ (corresponds to few microseconds with the expected non-radiative exciton lifetime exceeding nanosecond) in the slow-reservoir regime corresponding to $\Gamma/\gamma \sim 10^{2}$. 

\section{Broken radial symmetry} \label{sec:defect}

In stark contrast to predictions of reservoir-free models \cite{Nalitov2019}, the non-chiral burner-type condensates are stable within the more general model \eqref{eq:nonAd0Dmodel} even in the ideal circularly symmetric configuration.
Let us now test this property against the case of broken circular symmetry.
Although various mechanisms including planar disorder of the microcavity parameters can be responsible for the symmetry reduction, we only consider the one stemming from the imperfect optical pump shape.

Without loss of generality, we account for \textit{the azimuthal pump profile asymmetry} by adding a constant term $\delta W$ to the right-hand side of Eq.~\eqref{eq:dN1}.
This modification is expected to introduce an effective non-Hermitian potential grating and alter the allowed states of the condensate. 
In particular, the presence of the pump asymmetry  leaves only two burner-like states from the continuum, eliminating the ambiguity of the petals orientation.
The allowed states respect parity (mode swapping) symmetry, $S_x=\pm S$, and we refer to them as to symmetric ($S_x=S$) and anti-symmetric ($S_x=-S$) burners.
They are given by $S_y=S_z=N_2=0$, $N_1^\pm = \mp 2 N_0^\pm$, $S^\pm = (W-N_0^\pm)/(2N_\text{th})$ and 
\begin{eqnarray}
    N_0^\pm &=& \frac{1}{2}\left[3N_\textrm{th}+W - \right. \notag \\
    &-& \left.\sqrt{(3N_\textrm{th}+W)^2 - 4(W \mp \delta W) N_\textrm{th}}\right].
\end{eqnarray}

In contrast to symmetric case, the vortex-type solutions emerge only above a critical pumping $W_c \propto \delta W$ manifesting spontaneous breaking of the mode-swapping symmetry \cite{Nalitov2019,Chestnov2016,Chestnov2023}. 
Although the relevant $\bm{S}$-vectors are generally close to the $z$ axis, they also have nonzero in-plane components $S_x,S_y\ll|S_z|$ which generally can not be given in closed-form expressions.

\begin{figure}
    \centering
    \includegraphics[width = \linewidth]{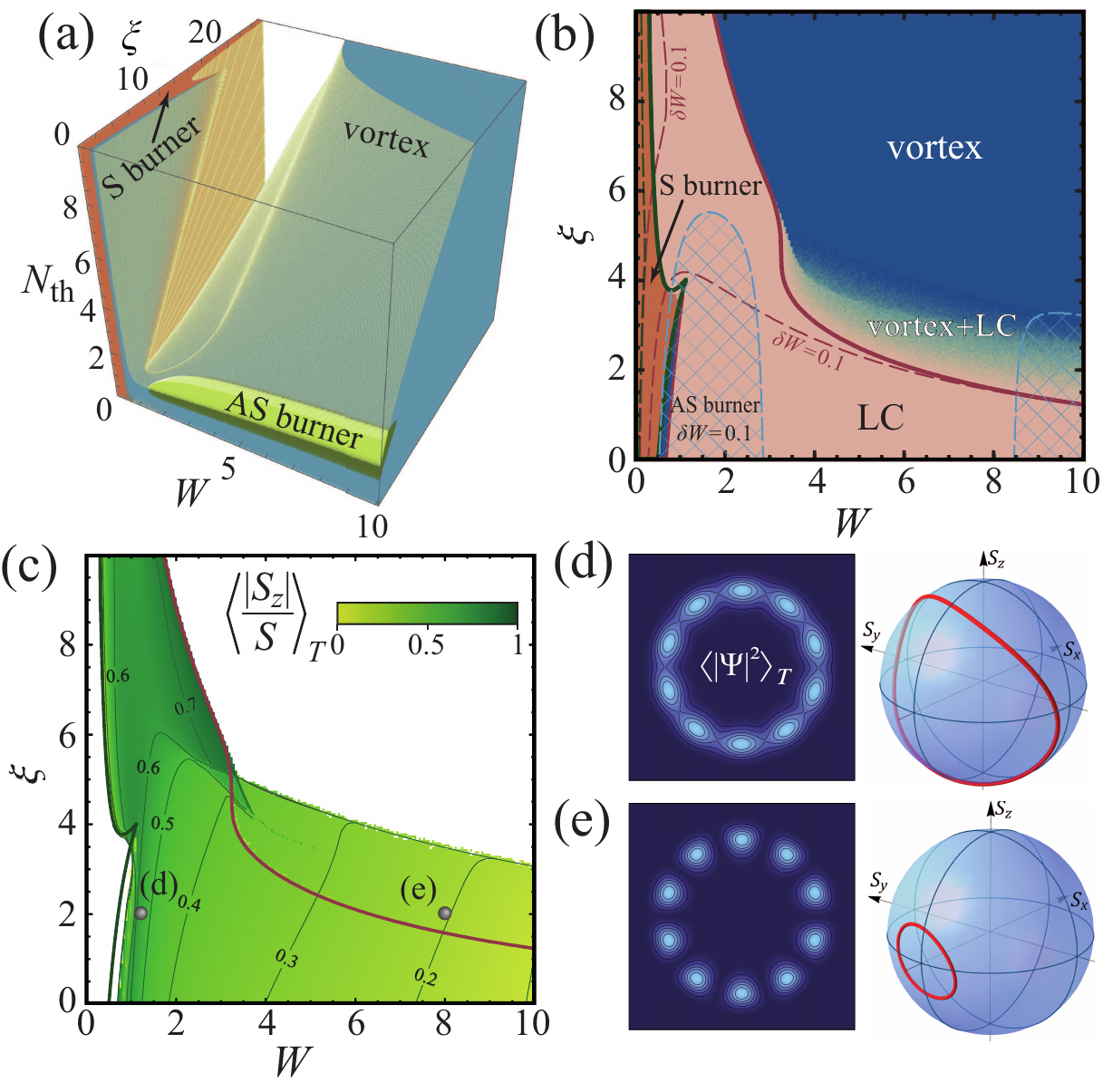}
    \caption{
    (a) Stability diagram at $\varepsilon=3$ and $\delta W=0.5$. The orange, yellow and blue regions correspond to stable symmetric burners, anti-symmetric burners and vortices, respectively.
    (b) Horizontal cross-section at $N_\text{th}=3$. The lines limit stability regions of vortices (red), symmetric (green) and anti-symmetric (blue) burners at $\delta W = 0.5$ (solid) and at $\delta W = 0.1$ (dashed). Realization probabilities of different states are shown with filling, blue for vortices, orange for symmetric burners (labeled as S burner), peach for limit cycles (LC). The hatched region corresponds to the stable anti-symmetric burners (labeled as AS burner) at $\delta W =0.1$.
    (c) Time-averaged distance of the direction vector $\bm{S}/S$ from the $(x,y)$-plane for the limit cycle trajectories from panel (b). 
    (d) and (e) Time averaged condensate density distribution at $l=5$ and the limit cycle loop in the classical spin space (projected into the unit sphere for clarity) for the two points from panel (c). Namely, at $\xi = 2$, $W=1.2$ (d) and at $\xi = 2$, $W=8$ (e).
    } 
    \label{fig:StAnalysis}
\end{figure}

Stability of numerically found vortex solutions and analytic burner-type states at the asymmetry strength $\delta W=0.5$ is shown in Fig.~\ref{fig:StAnalysis}(a). 
Here we consider the previously addressed case $\varepsilon=3$ to allow direct comparison.
Symmetric burners are stable only at the low pump powers close to the condensation threshold ($W\lesssim\delta W$), indicating that this solution type has a direct analogue in the reservoir-free models akin to Eq.~\eqref{eq:2mResFreeModel} \cite{Askitopoulos2018,Nalitov2019}.

Anti-symmetric burners, on the contrary, are stable at much stronger pumping $W\gg\delta W$.
Their stability region continuously evolves from that of self-trapping burners addressed in the previous section as the defect strength $\delta W$ increases. 
Note that such anti-symmetric burners are always unstable in the reservoir-free model \eqref{eq:2mResFreeModel} as it is shown, e.g., in Ref.~\cite{Nalitov2019}. 
This points out an intrinsically non-adiabatic nature of the stable anti-symmetric petalled condensates.

A horizontal cross-section of the 3D stability diagram at $N_\text{th}=3$ is demonstrated in Fig.~\ref{fig:StAnalysis}(b) which is an extension of Fig.~\ref{fig:Stability}(b) to the reduced symmetry case.
In addition to solid lines, which limit stability regions of vortices (red) and symmetric burners (green) at $\delta W=0.5$, the corresponding color-code dashed lines illustrate the weak-defect case $\delta W = 0.1$. 
The latter regime also features existence of the anti-symmetric burners (blue hatched region) whose stability domain is close to that of self-trapping states in the circularly symmetric case  shown in Fig. \ref{fig:Stability}(b).

Within the domains with no stable fixed-point solutions, the system evolves into the oscillating regime, see the peach domain in Fig.~\ref{fig:StAnalysis}(b). 
In addition, there is a region of bistability between the vortex and LC where the corresponding basins of attraction exhibit continuous redistribution as shown by the probability map in Fig.~\ref{fig:StAnalysis}(b).

Figures \ref{fig:StAnalysis}(c)--(e) further characterize numerically observed LC solutions. 
In particular, Fig.~\ref{fig:StAnalysis}(c) shows the time-averaged normalized population imbalance $\langle \left|S_z\right|/S \rangle_T$, where $\langle \cdot \rangle_T$ stands for the averaging over the period of LC oscillations.
This parameter describes how far the state deviates from the burner-type solution belonging to the $(x,y)$-plane.

The typical LC trajectories in the spin state projected onto the unit sphere and the corresponding time-averaged real-space polariton density distributions, assuming $l=5$, are shown in Figs.~\ref{fig:StAnalysis}(d) and (e).
It turns out that such oscillating states are hardly distinguishable from the stationary burner-like condensates under the typical experimental conditions with signal time-averaging.
Note that the limit cycle loop shown in Fig.~\ref{fig:StAnalysis}(d) qualitatively reproduces oscillations of the condensate vorticity reported in Ref.~\cite{Barrat2023}. 
The oscillation period $\sim 2$ [$\sim 7$ for Fig.~\ref{fig:StAnalysis}(e)] in real values corresponds to $~6 \Gamma^{-1}$ [$~21 \Gamma^{-1}$] in the considered case of $\Gamma/\gamma = 3$. 
For a typical polariton lifetime about few tens of picoseconds it results in the oscillation period exceeding hundred of picoseconds, in accordance with the results reported in Ref.~\cite{Barrat2023}.

\section{Comparison with adiabatic models}
\label{sec:adiabatic}

In previous sections, we pointed out the differences between predictions of the non-adiabatic two-mode model \eqref{eq:nonAd0Dmodel} and its reservoir-free symmetric approximation \eqref{eq:2mResFreeModel}.
However, the structure of the gain-saturation term in Eq.~\eqref{eq:2mResFreeModel}, originating from the condensate-to-reservoir coupling, stems from the linearization in the condensate density. 
This assumption is typically valid close to the condensation threshold only \cite{Yulin2023b}.
Thus the emerged discrepancies between the models could be attributed to linearization rather than non-adiabaticity.
Here we further investigate this aspect and compare stability diagrams in three models: 
\begin{enumerate}[itemsep=2pt,parsep=2pt,topsep=2pt, partopsep=2pt]
    \item the non-adiabatic model \eqref{eq:nonAd0Dmodel} developed in Sec.~\ref{sec:2modes};
    \item the model with adiabatically excluded reservoir;
    \item  the linearized adiabatic model \eqref{eq:2mResFreeModel} with included asymmetry.
\end{enumerate}

Let us first follow the reservoir elimination procedure \cite{bobrovska2015}.
The reservoir dynamics may be decoupled from the condensate evolution if its characteristic timescales are shorter than that of the condensate.
Under this assumption, the angular components of the quickly adopting reservoir density are expressed in the slowly varying condensate $\bm{S}$-vector from Eqs.~(\ref{eq:dN0}--\ref{eq:dN2}):
\begin{subequations}\label{eq:stationary_res}
    \begin{eqnarray}
        \label{eq:stationary_res_N0}
     N_0^{(0)} &=& { W(1+2S)-2N_\textrm{th}(S+S^2+S_z^2) - S_x\delta W \over 1+2S +2(S+S^2+S_z^2) }, \\
     N_1^{(0)} &=& {\delta W-2S_x(N_\textrm{th}+N_0^{(0)}) \over 1+2S}, \\ N_2^{(0)} &=& -{2S_y(N_\textrm{th}+N_0^{(0)}) \over 1+2S}, 
     \label{eq:stationary_res_N12}
    \end{eqnarray}
\end{subequations}
where we also account for the presence of pump asymmetry in Eq.~\eqref{eq:dN1}.

The deviations from the corresponding quasi-stationary values obey 
\begin{subequations}
    \begin{eqnarray}
     \dot{\delta N_0} &=& -(1+2S)\delta N_0-S_x \delta N_1 -S_y\delta N_2, \\
     \dot{\delta N_1} &=& -(1+2S)\delta N_1-2S_x \delta N_0, \\
     \dot{\delta N_2} &=& -(1+2S)\delta N_2-2S_y \delta N_0.
    \end{eqnarray}
\end{subequations}
Therefore, the evolution of perturbations is characterized by the relaxation rates $\gamma_0 = 1+2S$ and $\gamma_\pm=1+2S\pm\sqrt{2(S_x^2+S_y^2)}$ with the slowest rate being $\gamma_-$.

The general adiabatic model (ii) is obtained by substituting expressions \eqref{eq:stationary_res} into Eqs.~(\ref{eq:dSx}-\ref{eq:dSz}).
One can than extract the set of characteristic timescales of the fast condensate dynamics.
The terms originating from the Hermitian part of the corresponding Hamiltonian yield the self-induced Larmor precession about the $z$-axis  \cite{Shelykh2004}  with the frequency 
\begin{equation} \label{eq:LarmorFrequency}
    \Omega = \left( \xi - \varepsilon {N_\textrm{th}+N_0^{(0)} \over 1+2S} \right) S_z.
\end{equation}
The positive contribution $\xi S_z$ stems from the condensate self-repulsion while the negative one manifests effective attraction due to reservoir depletion.
In turn, the anti-Hermitian terms affecting the spin magnitude $S$ yield a couple of dissipation rates:
\begin{align}\label{eq:S_dissiapation_rates}
    \Gamma_z = N_0^{(0)}, \; \; \; \Gamma_{xy} = N_0^{(0)} - {N_\textrm{th} + N_0^{(0)} \over 1+2S}S ,
\end{align}
responsible for the decay of the out-of-plane $S_z$ and the in-plane $S_x$ and $S_y$ components, respectively.

The condensate dynamics can be self-consistently treated as slow  under the condition 
\begin{equation}
    \gamma_-\gg\Omega,\ \Gamma_z,\ \Gamma_{xy}.
\end{equation}

\begin{figure}
    \centering
    \includegraphics[width = \linewidth]{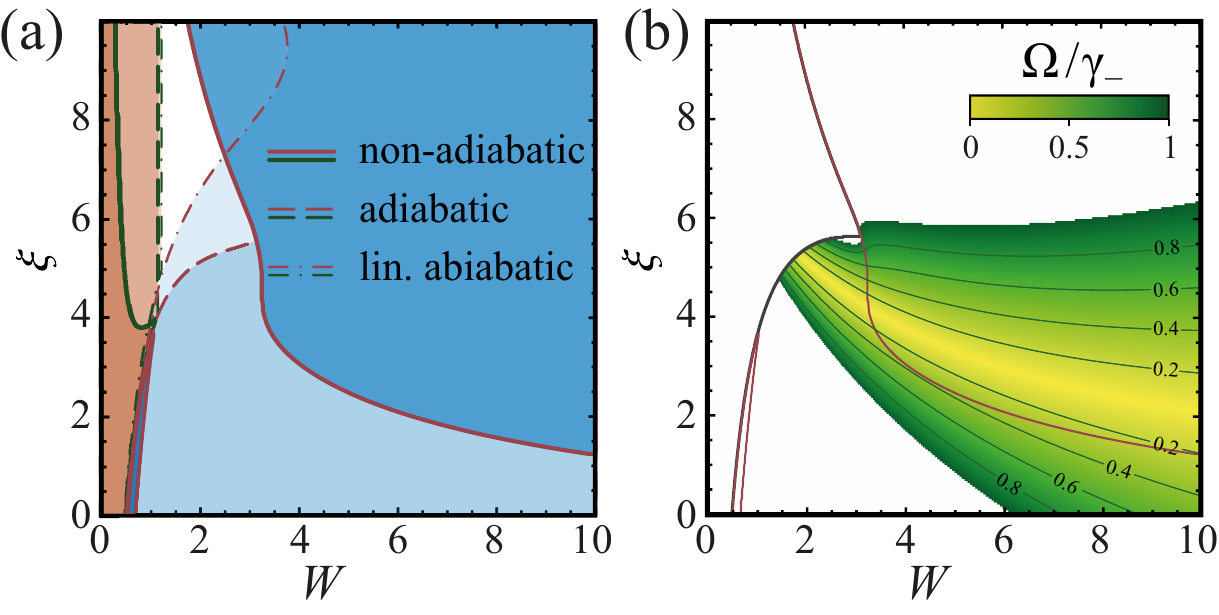}
    \caption{(a) Stability domains of the vortex (blue regions bounded by the red lines) and the burner (orange regions bounded by the green lines at $\varepsilon = 3$, $N_{\rm th}=3$ and $\delta W=0.5$. The solid lines correspond to the prediction of the full non-adiabatic model \eqref{eq:nonAd0Dmodel} which accounts for the presence of reservoir. The dashed lines correspond to the case of adiabatically excluded reservoir, while the dash-dotted lines correspond to the adiabatic model with linearized condensate-to-reservoir coupling Eqs.~\eqref{Eq:2modeModel}. 
    (b) The ratio between the absolute value of the effective Larmor precession frequency $\Omega$ and the slowest reservoir relaxation rate $\gamma_-$. The black line limits the vortex state existence domain while the red lines indicate the stability boundary shown on the panel (a).}
    \label{fig:5}
\end{figure}

The detailed comparison between the full (solid lines) and the adiabatic (dashed lines) models beyond the low-density approximation is demonstrated in Fig.~\ref{fig:5} for the same parameters as in Fig.~\ref{fig:StAnalysis}. 
The non-adiabatic behavior is manifested by the mismatch of the symmetric burner stability domains at large $\xi \gtrsim 4$, see the pale and bright orange shadings in Fig.~\ref{fig:5}(a). 
However, the most significant discrepancy occurs for the vortex solutions (blue shading) at the strong pumping and relatively weak self-repulsion $\xi$. 
Under these conditions, the slow condensate relaxation criterion $\Gamma_z,\ \Gamma_{xy} \ll \gamma_{-}$ still holds despite the fact that the considered case of $N_{\rm th}\equiv \Gamma/\gamma =3$ suggests the opposite \cite{bobrovska2015}. 
However the main reason of non-adiabatic behaviour is a large Larmor frequency $\Omega$, which benefits from the large out-of-plane $S_z$ component characteristic of the vortex state.

Indeed, the adiabaticity criterion $\Omega/\gamma_-\ll 1$ is satisfied only in a narrow region of the vortex existence domain, see Fig.~\ref{fig:5}(b).

Note that the analysis of the adiabatic model is severely complicated by an intricate structure of nonlinear terms given by Eqs.~\eqref{eq:stationary_res}.
That is why it is often reduced to its simplified analogue \eqref{eq:2mResFreeModel}. 
This requires linearization of the quasi-stationary reservoir density harmonics \eqref{eq:stationary_res}   in the $\bm{S}$-vector components assuming low condensate occupancy $S\ll1$.
This way, the asymmetric extension of model \eqref{eq:2mResFreeModel} can be derived from Eqs.~(\ref{eq:dSx})-(\ref{eq:dSz}):
\begin{subequations}\label{Eq:2modeModel}
    \begin{eqnarray}
        \dot{S}_x  &=& \left(W - 3 N_{\rm th} S  \right) S_x + \frac{\delta W}{2} S + \left(\xi - N_{\rm th}\varepsilon \right) S_y S_z, \label{eq:2modeModelSx}\\
        \dot{S}_y &=& \left(W - 3 N_{\rm th} S\right) S_y  - \frac{\varepsilon \delta W}{2} S_z  - \left(\xi  - N_{\rm th} \varepsilon \right) S_x S_z, \label{eq:2modeModelSy}\\
        \dot{S}_z  &=& \left(W - 2 N_{\rm th} S \right) S_y + \frac{\varepsilon \delta W}{2} S_y.
        \label{eq:2modeModelSz}
    \end{eqnarray}
\end{subequations}

Note that the pump asymmetry imprints an effective magnetic field $\varepsilon \delta W/2$ directed along $x$ and a gain imbalance term $\delta W/2$ in Eq.~\eqref{eq:2modeModelSx}. Both factors favour pinning of the state vector to the $x$-axis corresponding to the burner-type solution. 

According to the rigorous bifurcation analysis of Eqs.~\eqref{Eq:2modeModel} performed in \cite{Nalitov2019}, the symmetric burner is indeed the only stable solution at weak pumping $W\lesssim \delta W$.
However it inevitably loses stability above a certain critical pumping which depends on $\xi$ and $\delta W$, see the green dash-dotted line in Fig.~\ref{fig:5}(a).
In addition, at this or greater pumping [the red dash-dotted line in Fig.~\ref{fig:5}(a)], two opposite vortices appear which are \textit{always} the only stable solutions within the domain of their existence. 
Therefore, both simplified $\eqref{Eq:2modeModel}$ and non-linearized adiabatic models miss a wide vortex instability region predicted by the full model \eqref{eq:nonAd0Dmodel} (see the red solid and the red dash-dotted lines in Fig.~\ref{fig:5}(a)). 

In contrast to vortices, the anti-symmetric burners are always unstable within the linearized model \eqref{Eq:2modeModel}, see Ref.~\cite{Nalitov2019}. This result also holds for the general adiabatic model (ii) as we have checked numerically. It apparently contradicts with the non-adiabatic model \eqref{eq:nonAd0Dmodel} which supports stable self-trapping burners at the strong pumping and weak effective self-repulsion. 

\section{discussion and conclusions}

The simple reservoir-free models of polariton condensation in annular optical traps are able to explain formation of the standing-wave condensate with large angular number, in agreement with experimental results. However, within the adiabatic approximation, these states are doomed to be replaced by giant vortices as the pumping grows away from the condensation threshold. To the best of our knowledge, such a symmetry-breaking transition has not been observed in circular optical traps so far. We hypothesized it may be due non-adiabatic nature of polariton condensation.

Our analysis revealed several new phenomena connected with non-adiabatic behavior.
The most important features are manifested under the condition of elastic interaction with reservoir dominating inelastic scattering, $\varepsilon=\alpha/\beta>1$, and at weak self-repulsion $\xi$ where the reduced model predicts spontaneous vortex formation. 
In this regime, the stable non-chiral burner-shaped condensates exist far above the threshold even in the absence of external symmetry breaking.
The possible explanation of this phenomenon is the hole-burning effect \cite{Estrecho2018,Wouters2007a}. The burner-shaped condensate locally depletes reservoir giving it a form of a comb which acts as the effective potential grating. 
Under the conditions of slow reservoir, such a confining potential is rigid against weak perturbations in the condensate and thus is able to stabilize it.
In the case of reduced radial symmetry of the pump, these self-trapping solutions are transformed into a distinct type of anti-symmetric burners.

Another distinct manifestation of non-adiabatic behaviour is the instability of  vortex condensates at weak self-repulsion.  
Within this instability domain, the condensate experiences fast pulsations. 
The corresponding time-averaged condensate density distribution is practically hardly distinguishable from the burner states. 
Therefore, previous experimental studies may have overlooked this regime. 
A reliable detection of such fast oscillations requires time-delayed auto-correlation measurements as those performed in Refs.~\cite{Barrat2023,Kim2020,Sigurdsson2022}.

The stable vortex condensates are expected to appear either at $\varepsilon < 1$ or at $\varepsilon > 1$ in the regime of  strong pumping with large self-interaction parameter $\xi$. 
However the numerical computation of dimensionless self-repulsion form-factor $I_\xi$ shows that higher angular momentum condensates with $l \gg 1$, which are supported by wider traps, are characterised with lower values of $\xi$, see Fig.~\ref{fig:linear_model_b}.
This gives a possible interpretation to the absence of reported observations of such giant vortex condensates in annular optical traps.

In addition, at $\varepsilon<1$ and $\Gamma/\gamma\gg1$, a new self-induced oscillating regime is predicted, where the condensate is periodically switching between the two vortex states. 
These oscillations also exist in the traps with reduced symmetry as we have checked numerically.
Our estimates of the parameter $\varepsilon$ based on the available experimental data show that both $\varepsilon>1$ and $\varepsilon<1$ regimes are achievable and were in fact experimentally studied.

In conclusion, we mention other mechanisms which could potentially contribute to instability of vortex condensates on par with non-adiabaticity of the condensate-reservoir system.
Dynamic or dissipative interaction between the states with different angular momenta $|l|$, allowed by the selection rules of effective photonic spin-orbit coupling \cite{Yulin2020}, could result in bistability or limit cycle behaviour involving different orbital momentum states.
Furthermore, interaction-induced mixing of radial modes, which is neglected in our model, eventually leads to the interaction-dominated Thomas-Fermi regime of condensate, where vortices with cores of the size of the healing length are formed, at the pumping powers well above the condensation threshold \cite{Keeling2008}.

Instability of multiply charged optically trapped polariton vortices, $l \gg 1$, attributed to Kelvin-Helmholtz mechanism was also observed in numerical simulations in the strongly nonlinear regime well above condensation threshold \cite{Alperin2021}.
In this case, however, instability switches on once the rotation speed reaches the local speed of sound $v_c \propto \sqrt{|\Psi|^2}$.
On the contrary, in the vicinity of the condensation threshold the vortex speed is governed by trap size and the nonadiabaticity-induced instability occurs once the condensate density $|\Psi|^2\propto s \propto W$ reaches a critical value (see Fig. \ref{fig:Stability}).

In the intermediate range of pumping powers above the condensation threshold, formation of periodically evolving vortex-antivortex clusters with spatially separated cores was reported in \cite{Sitnik2022}.
As the orientation of clusters was pinned, this periodic behaviour was attributed to beats of optically trapped polariton modes.
However, no spontaneous symmetry breaking with formation of a persistent rotating polariton condensate was observed, indicating instability of such states with fixed vorticity sign, which finds interpretation in our work.

\section{acknowledgements}

We thank Alexey Yulin and Stanislav Baturin for valuable discussions.
The work of A.N. (analytical calculations, elaboration of the model) is supported by he Russian Science Foundation under Grant No. 22-12-00144. 
The work of I.C. (numerical analysis of the non-Hermitian properties) is supported by the Russian Science Foundation Grant No. 22-72-00061. 
E.C. (analytical calculations) acknowledges the support from Basis foundation, Grant No.21-1-3-30-1.

\appendix 
\section{Derivation of the non-adiabatic two-mode model of polariton condensation} \label{AppA}

In this section we provide a derivation of the dimensionless two-mode model of non-adiabatic polariton condensation in the ring-shaped optical traps. We start from the general two-dimensional model \eqref{eq:FullModel} and account for the delta-ring shape of the pump, $P(r) = P_0 \delta(r-R)/R$. Using the two-mode representation of the condensate wave function \eqref{eq:WF} and the corresponding definition \eqref{eq:Res_Den_Definition} for the reservoir density, we integrate out the radial and the azimuthal degrees of freedom. The resulting equation for the total condensate occupancy parameterized by the magnitude of the classical spin $\bm{s}=\psi^\dagger \bm{\sigma} \psi/2$ then reads:
\begin{equation}
\dot{s} = \pi b \left[ 2 n_0 s + n_1 s_x + n_2 s_y \right]. \label{eq:ds}
\end{equation}
Here $b=\beta|\Psi_l(R)|^2 /\hbar$ arises from the overlap between the condensate and the thin-ring reservoir which follows the shape of the pump. 

The spin components, in turn, obey the following equations:
\begin{subequations}\label{eq:s_and_n}
\begin{eqnarray}
    \dot{s}_x &=& \pi b \left[ 2 n_0 s_x + n_1 s + \varepsilon n_2 s_z \right] + a s_y s_z, \label{eq:dsx}\\
    \dot{s}_y &=& \pi b \left[ 2 n_0 s_y + n_2 s - \varepsilon n_1 s_z \right] -a s_x s_z, \label{eq:dsy}\\
    \dot{s}_z &=& \pi b \left[ 2 n_0 s_z + \varepsilon (n_1 s_y - n_2 s_x) \right], \label{eq:dsz}
\end{eqnarray}
where $a = 4\pi \alpha_1 \int_0^\infty |\Psi_l|^4 r dr / \hbar$ describes the strength of the condensate self-interactions accounting for its density profile, and we defined $\varepsilon=\alpha/\beta$.

The equations for the corresponding reservoir density components \eqref{eq:Res_Den_Definition} read:
\begin{eqnarray}
    \dot{n_0} &=& P - P_\textrm{th} - \gamma n_0 - 2bs(n_\textrm{th}+n_0) -
    \\ \notag &&  b(s_xn_1+s_yn_2),\label{eq:dn0}\\
    \dot{n_1} &=& - (\gamma + 2bs) n_1 - 2bs_x(n_\textrm{th}+n_0), \label{eq:dn1}\\
    \dot{n_2} &=& - (\gamma + 2bs) n_2 - 2bs_y(n_\textrm{th}+n_0), \label{eq:dn2}
\end{eqnarray}
\end{subequations}
where $n_\textrm{th} = P_\textrm{th}^{l}/\gamma$ is the reservoir density at the threshold pumping $P_0 = P_\textrm{th}^{l}$ corresponding to condensation in the vortex state with azimuthal number $l$. 
Note that the value of $l$ is not arbitrary but is uniquely determined by the linear parameters of the problem, see Sec.~\ref{subsec:IIA}.

Then we introduce the dimensionless time $\tau = \gamma t$ and use the dimensionless representations of the spin vector $\bm{S} = b\bm{s}/\gamma$ and the reservoir density components $N_j = 2\pi b n_j/\gamma$. 
This yields the dimensionless version of Eqs.~\eqref{eq:s_and_n} which corresponds to Eqs.~\eqref{eq:nonAd0Dmodel} in the main text. 
With these notations, the threshold density of the reservoir reads $N_\textrm{th} = 2\pi bn_\textrm{th}/ \gamma$. 
Here the numerator stands for the effective gain rate feeding the state $l$. In the threshold conditions, this gain is compensated by the dissipation $\Gamma$ term.  
Therefore, we obtain $N_\textrm{th} = \Gamma / \gamma$. Below we will rigorously derive this expression from the condensate continuity condition.

For this purpose we resort to the stationary version of one-dimensional Gross-Pitaevskii equation (GPE) which governs the radial distribution of the condensate rotating with the angular quantum number $l$. 
At the threshold pumping rate, the reservoir remains in the steady state with a constant density $N_{\rm th}$.
Therefore, the radial part of the condensate wave function obeys:
\begin{align}\label{Eq.A1}
  \left[  - \frac{1}{\rho}\partial_{\rho} \left( \rho \partial_{\rho} \right)
   + \frac{l^2}{\rho^2} - i + (\varepsilon + i) p_{\rm th}^l \frac{\delta(\rho - \rho_0)}{\rho_0} \right]\Psi_l(\rho) = \\
    = \frac{2\omega_0}{\Gamma}\Psi_l(\rho), \nonumber
\end{align}
where $\rho = r\sqrt{m\Gamma/\hbar}$ is the normalized radial coordinate and $\rho_0$ stands for the normalized pump radius. At the threshold pumping strength $p_{\rm th}^l=N_{\rm th} \beta m \gamma / \left( 2\pi b \hbar^2  \right) =P_{\rm th}^l\beta m/(\hbar^2\gamma)$ the condensate frequency $\omega_0$ is real. 
Using Eq.~\eqref{Eq.A1} it is straightforward to obtain an effective continuity equation which reads at the threshold:
\begin{equation} \label{eq:continuity}
   \rho^{-1}\partial_\rho\left(\rho j\right)= \left(p^{l}_{\rm th}\frac{\delta(\rho - \rho_0)}{\rho_0} -1\right) \left|\Psi_l(\rho)\right|^2,
\end{equation}
where $j = \left(\Psi^* \partial_\rho \Psi- \Psi \partial_\rho \Psi^*\right) / \left(2i\right)$ stands for the dimensionless radial current. 
The right-hand side of Eq.~\eqref{eq:continuity} stands for the  inflow of polaritons from the outer region while the left-hand side describes the local gain-dissipation balance.
Integrating the last expression over radial coordinate yields
\begin{equation} \label{Eq.A3}
    p^{l}_{\rm th}  \left|\Psi(\rho_0)\right|^2 = {m\Gamma \over {2\pi \hbar} }+ \left(\rho j \right)\biggr|_{0}^{\infty},
\end{equation}
where we account for the normalization condition $\int \left|\Psi_l\right|^2 \rho d\rho = {m\Gamma \left/ \left( {2\pi \hbar} \right)\right.}$. 
The last term in \eqref{Eq.A3} vanishes since $\lim_{\rho\to 0} j \neq \infty$ and $\lim_{\rho\to\infty} j = 0$. 
Therefore, the threshold pumping strength which governs the net gain rate is governed only by the condensate density at the pump position:
\begin{equation} \label{eq:threshold}
    p^{l}_{\rm th} = \frac{m\Gamma}{2\pi \hbar \left|\Psi(\rho_0)\right|^{2}}.
\end{equation}
Using expression \eqref{eq:resonance} for the wave function modulus we obtain:
\begin{equation}\label{eq.A5}
    p_{\rm th}^l =  \left[ \int_0^{\rho_0}\left|J_l(\kappa\rho) \over J_l(\kappa \rho_0) \right|^2 \rho d \rho + \int_{\rho_0}^\infty\left|H_l(\kappa\rho) \over H_l(\kappa \rho_0) \right|^2 \rho d \rho\right].
\end{equation}

This expression implicitly connects the threshold pumping strength with the pump radius $\rho_0$ through the condensate frequency $\omega_0$ which enters the parameter $\kappa=\sqrt{2\omega_0/\Gamma + i}$ governed by the eigenvalue problem \eqref{Eq.EigenProblem}.

According to the definition, $\xi = a/b$, where $a$ and $b$ quantifies the overlap of the condensate with itself and with the reservoir, respectively. 
Since we consider the limit of a $\delta$-shaped reservoir, the $b$-parameter is nothing but the condensate gain rate at the pump position, $b = \left. \beta \left|\Psi(\rho_0)\right|^2\right/ \hbar$. 
Therefore, using Eq.~\eqref{eq:threshold} and calculating parameter $a$ with the wave function definition \eqref{eq:resonance} we obtain $\xi = I_{\xi}\alpha_1/\beta$. 
Here the form factor  $I_{\xi}$ is connected with dimensionless threshold parameter:
\begin{equation} \label{eq:Ixi_App}
    I_\xi = \frac{2}{p_\textrm{th}^l} \left[ \int_0^{\rho_0} \left|\frac{J_l(\kappa\rho)}{J_l(\kappa\rho_0)}\right|^4 \rho d\rho + \int_{\rho_0}^{\infty} \left|\frac{H_l(\kappa\rho)}{H_l(\kappa\rho_0)}\right|^4 \rho d\rho   \right],
\end{equation}
see Eq.~\eqref{eq:Ixi}.

Note that Eq.~\eqref{eq:threshold} can be represented as $p^{l}_{\rm th} = \gamma \Gamma /\left(2\pi b\right) \cdot \beta m /\left( \hbar^2 \gamma \right)$. Combining it with the expression for $p^{l}_{\rm th}$ given after Eq.~\eqref{Eq.A1} we finally obtain $N_{\rm th} = \Gamma/\gamma$ which manifests a balance between the net gain and the losses stated in the continuity equation \eqref{eq:continuity}.

\bibliography{references}

\end{document}